\documentclass[12pt]{elsart}
\usepackage{feynmp,epsfig,graphics}
\def\wslash{\FMslash w}

\def\prt{\partial}

\begin{document}
GSI-Preprint-2003-17
\begin{frontmatter}
\title{On baryon resonances and chiral symmetry }
\author[NBI]{E.E. Kolomeitsev}
\author[GSI,TU]{and M.F.M. Lutz}
\address[NBI]{The Niels Bohr Institute\\ Blegdamsvej 17, DK-2100 Copenhagen\\Denmark}
\address[GSI]{Gesellschaft f\"ur Schwerionenforschung (GSI),\\
Planck Str. 1, 64291 Darmstadt, Germany}
\address[TU]{Institut f\"ur Kernphysik, TU Darmstadt\\
D-64289 Darmstadt, Germany}
%\today
\begin{abstract}
We study $J^P=(\frac{3}{2})^-$ baryon resonances as  generated by
chiral coupled-channel dynamics in the $\chi-$BS(3) approach.
Parameter-free results are obtained in terms of the Weinberg-Tomozawa term
predicting the leading s-wave interaction strength of Goldstone bosons with
baryon-decuplet states. In the 'heavy' SU(3) limit with $m_\pi =
m_K \sim 500 $ MeV the resonances turn into bound states forming a
decuplet and octet representation of the SU(3) group. Using physical
masses the mass splitting are remarkably close to the empirical pattern.
\end{abstract}
\end{frontmatter}
%\tableofcontents

\section{Introduction}

In this letter we further test the conjecture
\cite{Wyld,Dalitz,Ball,Rajasekaran,Wyld2,LK01,LK02,LWF02,LH02}
that baryon resonances not belonging to the large-$N_c$
ground states may be generated dynamically by coupled-channel
dynamics. In recent works \cite{LK01,LK02,LK00,Granada} it was
shown that chiral dynamics as implemented by the $\chi-$BS(3) approach
\cite{LK01,LK02,LH02} provides a parameter-free prediction for the
existence of a wealth of s-wave baryon resonances. The latter may be classified in the 'heavy'
SU(3) limit as forming two mass-degenerate octet and one
singlet states \cite{Wyld,Dalitz,Ball,Rajasekaran,Wyld2,Granada,Jido03}.
Related works that adhere to a different strategy not insisting on improved
crossing-transformation properties of the scattering amplitudes are
\cite{grnpi,grkl,Oset-prl,Oset-plb,Jido03}. All these findings
support our conjecture. In the SU(6) quark-model approach such s-wave resonances belong to a
$70$-plet, that contains many more resonance states \cite{Schat}. An
interesting question arises: what is the role played by the
d-wave resonances belonging to the very same $70$-plet as the
s-wave resonances studied in \cite{Granada}. The phenomenological
model \cite{LWF02} generated successfully also non-strange d-wave
resonances belonging to the $70$-plet by coupled-channel dynamics
describing a large body of pion and photon scattering data.

Naively one may expect that chiral dynamics does not make firm
predictions for d-wave resonances since the meson-baryon
interaction in the relevant channels probes a set of
counter terms presently unknown. However, this is not
necessarily so. Since a d-wave baryon resonance couples to s-wave
meson baryon-decuplet states chiral symmetry is quite predictive
for such resonances under the assumption that the
latter channels are dominant. This is in full analogy to the
analysis of the s-wave resonances  \cite{Wyld,Dalitz,Ball,Rajasekaran,Wyld2,LK00,Granada,Jido03} that neglects
the effect of the contribution of d-wave meson baryon-decuplet
states. The empirical observation that the d-wave resonances
$N(1520)$, $N(1700)$ and $\Delta (1700)$ have large branching
fractions ($> 50 \% $) into the inelastic $N \pi \pi$ channel, even
though the elastic $\pi N$ channel is favored by phase space,
supports our assumption. A parameter-free scheme arises since the
Weinberg-Tomozawa theorem predicts the leading s-wave interaction
strength of Goldstone bosons not only with baryon-octet but also with
baryon-decuplet states.

In this letter we explore whether the Weinberg-Tomozawa
interaction of the meson baryon-decuplet sector has the potential
to dynamically generate d-wave baryon resonances. We find that the
chiral dynamics predicts the existence of octet and decuplet
bound states in the 'heavy' SU(3) limit. Those bound states disappear
once the current quark masses of QCD are sufficiently small.
It should be possible to test this prediction within lattice QCD~\cite{dgr}. Using physical
current quark masses the predicted mass spectrum is remarkably consistent with the empirical
spectrum. This finding complements a corresponding result
\cite{Wyld,Dalitz,Ball,Rajasekaran,Wyld2,Granada,Jido03} obtained
for the s-wave baryon resonances for which chiral dynamics predicts 2 degenerate octet
and one singlet state that also disappear for sufficiently small current quark masses.

\section{Chiral coupled-channel dynamics: the $\chi$-BS(3) approach}

The starting point is the chiral SU(3) Lagrangian (see e.g.\cite{Krause}). The relevant
term that encodes the prediction of Weinberg and Tomozawa \cite{Wein-Tomo} for
the s-wave scattering lengths of Goldstone bosons with baryon-decuplet states is
readily identified,
\begin{eqnarray}
\mathcal{L}_{WT}= \frac{3\,i}{8\, f^2}\, {\rm tr}\, \Big((\bar B_\nu
\,\gamma^\mu\,\, B^\nu) \cdot
 [\Phi,(\prt_\mu \Phi)]_-  \Big)\,,
 \label{WT-term}
\end{eqnarray}
where we dropped terms that do not contribute to the
on-shell scattering process at tree level. The SU(3) meson and
baryon fields are written in terms of their isospin symmetric
components,
\begin{eqnarray}
\Phi &=& \tau \cdot  \pi + \alpha^\dagger \!\cdot \! K + K^\dagger
\cdot \alpha  + \eta \,\lambda_8 \;,
\nonumber\\
\bar B \cdot B &=& \frac{1}{6}\,\Big( \bar \Delta \,{\mathcal G
}\, \Delta \Big)\,\cdot \tau +\frac{1}{6}\,\bar \Delta \cdot
\Delta \,\Big( 2+\sqrt{3}\,\lambda_8 \Big) +\frac{1}{3}\,\bar
\Sigma \cdot \Sigma -\frac{i}{3}\, \tau \cdot \Big( \bar \Sigma
\times \Sigma \Big)
\nonumber\\
&+& \frac{1}{6}\,\bar \Xi \,\Xi \,\Big( 2-\sqrt{3}\,\lambda_8\Big)
+\frac{1}{6}\,\Big(\bar \Xi \,\sigma \, \Xi \Big)\,\tau +
\frac{1}{3}\,\bar \Omega^-\,\Omega^- \,\Big(1-\sqrt{3}\,\lambda_8
\Big)
\nonumber\\
&+&\frac{1}{\sqrt{6}}\,\bar \Delta \,\Big(\Sigma \cdot
T\Big)\,\alpha + \frac{1}{\sqrt{6}}\, \alpha^\dagger \,\Big(\bar
\Sigma \cdot T^\dagger \Big)\,\Delta -\frac{1}{3}\, \bar \Sigma \,
\Big( \Xi^t \,i\,\sigma_2\,\sigma \,\alpha  \Big)
\nonumber\\
&+& \frac{1}{3}\, \Big( \alpha^\dagger \,\sigma
\,i\,\sigma_2\,\bar \Xi^t\Big) \,\Sigma + \frac{1}{\sqrt{6}}\,\bar
\Omega^-\,\Big(\alpha^\dagger \cdot \Xi \Big)
+\frac{1}{\sqrt{6}}\, \Big(\bar \Xi \cdot \alpha \Big)\,
\Omega^-\,, \, \label{field-decomp}
\end{eqnarray}
with the  Gell-Mann matrices, $\lambda_i$, and the isospin doublet fields
$K =(K^+,K^0)^t $ and $\Xi = (\Xi^0,\Xi^-)^t$.
The isospin Pauli matrices $\sigma=(\sigma_1,\sigma_2,\sigma_3)$ act
exclusively in the space of isospin doublet fields $(K,N,\Xi)$ and
the matrix valued isospin doublet $\alpha$,
\begin{eqnarray}
&& \alpha^\dagger =
 {\textstyle{1\over\sqrt{2}}}\left( \lambda_4+i\,\lambda_5 ,
\lambda_6+i\,\lambda_7 \right) \;,\;\;\;\tau =
(\lambda_1,\lambda_2,\lambda_3)\;, \nonumber\\
&& {\mathcal G}_i = 3\,\vec T\,\sigma_i\,\vec T^\dagger \,,\qquad
\vec T \cdot \vec T^\dagger =1 \,, \qquad
T^\dag_i \, T_j=\delta_{ij}-{\textstyle{1\over 3}}\,\sigma_i\,\sigma_j\,.
 \label{def-alpha}
\end{eqnarray}
The 4$\times$2 matrices $T_j$ in (\ref{def-alpha}) describe the
transition from isospin-$\frac{1}{2}$ to isospin-$\frac{3}{2}$ states.
The scattering process is described by the amplitudes that follow as solutions of the
coupled-channel Bethe-Salpeter equation,
\begin{eqnarray}
T_{\mu \nu}(\bar k ,k ;w ) &=& K_{\mu \nu}(\bar k ,k ;w )
+\int\!\! \frac{d^4l}{(2\pi)^4}\,K_{\mu \alpha}(\bar k ,
l;w )\, G_{\alpha \beta}(l;w)\,T_{\beta \nu}(l,k;w )\;,
\nonumber\\
G_{\mu \nu }(l;w)&=&-i\,D(\half\,w-l)\,S_{\mu \nu}( \half\,w+l)\,,
\label{BS-coupled}
\end{eqnarray}
where we suppress the coupled-channel structure for simplicity. The meson and
decuplet propagators,  $D(q)$ and $S_{\mu \nu}(p)$, are used in the notation of \cite{LWF02}.
The scattering amplitude $T_{\mu \nu}(\bar k,k;w)$ decouples into various sectors
characterized by isospin (I) and strangeness (S) quantum numbers. This decomposition
follows from a corresponding one (see \cite{LWF02}) of the
interaction Lagrangian,
\begin{eqnarray}
{\mathcal L}_{WT}(\bar k ,k ;w)&=&
\sum_{I,S}\,R_\mu^{(I,S)\,\dagger }(\bar q,\bar p)\,\gamma_0\,
\,g^{\mu \alpha}
\,K^{(I,S)}_{\alpha \beta}(\bar k ,k ;w )\,g^{\beta \nu}\,R^{(I,S)}_\nu(q,p) \,,\;\;\;\;
\label{}
\end{eqnarray}
where the column $R^{(I,S)}(q,p)$ specifies our phase convention for the isospin states.
We introduced convenient kinematics:
\begin{eqnarray}
w = p+q = \bar p+\bar q\,,
\quad k= \half\,(p-q)\,,\quad
\bar k =\half\,(\bar p-\bar q)\,,
\label{def-moment}
\end{eqnarray}
where $q,\,p,\, \bar q, \,\bar p$ are the initial and final meson and baryon 4-momenta.
In Tab. \ref{tab:states} we collect $R^{(I,S)}(q,p)$ for all isospin and strangeness channels considered
in this work. Following the $\chi$-BS(3) approach developed in \cite{LK02,LWF02} the
interaction kernel is decomposed into a set of covariant projectors that have well defined
total angular momentum ($J$), and parity ($P$),
\begin{eqnarray}
&& K_{\mu \nu}(\bar k ,k ;w )  = \sum_{J,P}\,K^{(J,P)}(\sqrt{s}\,)\,
{\mathcal Y}^{(J,P)}_{\mu \nu}(\bar q, q,w) \,,
\nonumber\\
&& {\mathcal Y}^{(1/2, \pm)}_{\mu \nu}(\bar q,q;w) = \frac{1}{6}
\left(\gamma_\mu -\frac{w_\mu}{w^2}\,\wslash \right)
\left(\mp \,1 - \frac{\wslash }{\sqrt{w^2}}\right)
\left(\gamma_\nu -\frac{w_\nu}{w^2}\,\wslash \right) \;.
\nonumber\\
&& {\mathcal Y}^{(3/2, \pm)}_{\mu \nu}(\bar q,q;w) = \frac{1}{2}\left( g_{\mu \nu}-\frac{w_\mu
\,w_\nu}{w^2}\right)
\left(\mp \,1 + \frac{\wslash }{\sqrt{w^2}}\right)
\nonumber\\
&& \qquad \qquad \qquad \quad -\frac{1}{6}
\left(\gamma_\mu -\frac{w_\mu}{w^2}\,\wslash \right)
\left(\mp \,1 - \frac{\wslash }{\sqrt{w^2}}\right)
\left(\gamma_\nu -\frac{w_\nu}{w^2}\,\wslash \right) \;.
\label{def-proj}
\end{eqnarray}
where we provide the projector, ${\mathcal Y}^{(1/2,\pm)}_{\mu \nu}(\bar q,q;w)$ describing p- and d-wave scattering
with $J=1/2$ and ${\mathcal Y}^{(3/2,\pm)}_{\mu \nu}(\bar q,q;w)$ relevant for s-and p-wave scattering with
$J=3/2$. The merit of the projectors is that they decouple the Bethe-Salpeter
equation (\ref{BS-coupled})
into orthogonal sectors labelled by the total angular momentum $J$. Here we suppress
an additional matrix structure that follows since for given parity and total angular momentum,
$J \geq 3/2$, two distinct angular momentum states couple. In general, for given $J$, the
projector form a $2\times 2$ matrix, for which we displayed in (\ref{def-proj}) only its
leading $11$-component. The effect of the remaining components is phase-space suppressed and
not considered here. Referring to the detailed discussion given in \cite{LK02} we assume a
systematic on-shell reduction of an effective interaction kernel, $$V =K+ {\mathcal O}(Q^3)\,,$$
which is expanded according to chiral power counting rules. In addition, we
insist on the renormalization condition,
\begin{eqnarray}
T_{\mu \nu}^{(I,S)}(\bar k,k;w)\Big|_{\sqrt{s}= \mu (I,S)} =
V_{\mu \nu}^{(I,S)}(\bar k,k;w)\Big|_{\sqrt{s}= \mu (I,S)} \,,
\label{ren-cond}
\end{eqnarray}
that complies with approximate crossing symmetry. For the subtraction points,
$\mu(I,S)$, the natural choices are determined by the baryon-octet masses,
\begin{eqnarray}
&& \mu(I,+1)=\mu(I,-3)={\textstyle{1\over 2}}\,(m_\Lambda+ m_\Sigma) \,,
\quad \mu(I,0)=m_N\,, \quad
\nonumber\\
&&  \mu(0,-1)=m_\Lambda,\quad \mu(1,-1)=m_\Sigma\,, \quad
\mu(I,-2)= \mu(I,-4)= m_\Xi
 \label{eq:sub-choice}
\end{eqnarray}
as explained in detail in \cite{LK02}. The renormalization condition reflects
the fact that at subthreshold energies the scattering amplitudes may be
evaluated in standard chiral perturbation theory with the typical
expansion parameter $m_K/(4 \,\pi f) < 1 $ with $f \simeq 90$
MeV. Once the available energy is sufficiently high to permit elastic
two-body scattering a further typical dimensionless parameter
$m_K^2/(8\,\pi f^2) \sim 1$ arises. Since this ratio is uniquely
linked to two-particle reducible diagrams it is sufficient to sum
those diagrams keeping the perturbative expansion of all irreducible
diagrams. This is achieved by (\ref{BS-coupled}).  The subtraction points
(\ref{eq:sub-choice}) are the unique choices that protect the s-channel
baryon-octet masses manifestly in the p-wave $J={\textstyle{1\over 2}}$
scattering amplitudes.  The merit of the scheme \cite{LK00,LK01,LK02} lies in the
property that for instance the $K \,\Xi$ and $\bar K\,\Xi $
scattering amplitudes match at $\sqrt{s} \sim m_\Xi $
approximately as expected from crossing symmetry. The subtraction points
(\ref{eq:sub-choice}) can also be derived if one incorporates photon-baryon
inelastic channels. Then additional crossing symmetry constraints
arise. For instance the reaction $\gamma \,\Xi \to \gamma \,\Xi $,
which is subject to a crossing symmetry constraint at threshold, may
go via the intermediate states $\bar K \,\Lambda $ or $\bar K \,\Sigma $.
Here we assume that this reaction is described by a coupled-channel
scattering equation (\ref{BS-coupled})
where the effective on-shell interaction kernel $V$ is expanded in
chiral perturbation theory.

\begin{table}
\tabcolsep=-.1mm
\begin{tabular}{||c|c|c|c||}
\hline\hline
$(\frac12,-4)$ &
$(0,-3)$ &
$(1,-3)$ &
$(\frac12, -2)$
\\\hline
$(\overline{K}\,\Omega^\mu)$ &
$\left(\begin{array}{c}
({\textstyle{1\over \sqrt{2}}}\,\overline{K}\,\Xi^\mu)\\ (\eta\, \Omega^\mu)
\end{array}\right)$ &
$\left(\begin{array}{c}
(\pi\,\Omega^\mu)\\({\textstyle{1\over \sqrt{2}}}\,\overline{K}\,\sigma\,\Xi^\mu)
\end{array}\right)$ &
$\left(\begin{array}{c}
({\textstyle{1\over \sqrt{3}}}\,\pi \cdot \sigma \,\Xi^\mu)\\
({\textstyle{i\over \sqrt{3}}}\,\Sigma^\mu \cdot \sigma \,\sigma_2\,\overline{K}^t )\\
(\eta\,\Xi^\mu)\\(K\,\Omega^\mu) \end{array}\right)$
\\\hline\hline
$(\frac32,-2)$ &
 $(0,-1)$ &
 $(1,-1)$  &
 $(2,-1)$
 \\ \hline
$\left(\!\!\begin{array}{c}( \pi \cdot T\,\Xi^\mu )\\  (\Sigma^\mu \cdot T\, i\sigma_2
\overline{K}^t)\end{array}\!\!\right)$ &

$\left(\!\!\begin{array}{c}({\textstyle{1\over \sqrt{3}}}\,\pi\cdot  \Sigma^\mu)\\
({\textstyle{1\over \sqrt{2}}}\, K^t\,i\sigma_2\Xi^\mu)\end{array}\!\!\right)$&

$\left(\!\!\begin{array}{c}({\textstyle{-i\over \sqrt{2}}}\,\pi\times
\Sigma^\mu)\\( \sqrt{{\textstyle{3\over 4}}}\,\overline{K}\,\vec T^\dagger\, \Delta^\mu)
\\(\eta\, \Sigma^\mu)\\
({\textstyle{1\over \sqrt{2}}}\,K^t\,i\sigma_2\sigma \Xi^\mu)\end{array}\!\!\right)$ &

$\left(\!\!\begin{array}{c}
{\textstyle{1\over 2}}(\pi_i \,\Sigma^\mu_j+\pi_j \,\Sigma^\mu_i) -{\textstyle{1\over 3}}\, \delta_{ij}\,\pi \cdot
\Sigma^\mu\\
{\textstyle{1\over \sqrt{8}}}\,\overline{K}\,(\sigma_i\,T^\dagger_j + \sigma_j\,T_i^\dagger )\,\Delta^\mu
\end{array}\!\!\right)$
\\\hline\hline
 $(\frac12, 0)$&
 $(\frac32, 0)$ &
\multicolumn{2}{|c||}{$(\frac52,0)$}
\\ \hline
$\left(\begin{array}{c}({\textstyle{1\over \sqrt{2}}}\,\pi\cdot T^\dagger \,
\Delta^\mu)\\({\textstyle{1\over \sqrt{3}}}\,\Sigma^\mu\cdot \sigma \, K )\end{array}\right)$ &
$\left(\begin{array}{c}({\textstyle{1\over \sqrt{15}}}\,\pi \cdot {\mathcal G}\, \Delta^\mu)\\
(\eta\,\Delta^\mu)\\(\Sigma^\mu\,T\,K)\end{array}\right)$ &
\multicolumn{2}{|c||}{
$\begin{array}{c}\Big(( {\textstyle{1\over 2}}\,(\pi_i\,T^\dagger_j + \pi_j\,T^\dagger_i)
- {\textstyle{1\over 3}}\,\delta_{ij}\,\pi \cdot T^\dagger )\,\Delta^\mu\Big)
\end{array}$}
\\
\hline\hline
\multicolumn{2}{||c|}{$(1,1)$} &
\multicolumn{2}{|c||}{$(2,1)$}
\\ \hline
\multicolumn{2}{||c|}{$ ( \sqrt{{\textstyle{3\over 4}}}\,K^t\,i\,\sigma_2\,\vec T^\dagger \, \Delta^\mu)$ }
&
\multicolumn{2}{|c||}{
$ ( {\textstyle{1\over \sqrt{8}}}\,K^t\,i\,\sigma_2\,
(\sigma_i\,T^\dagger_j + \sigma_j\,T_i^\dagger
)\,\Delta^\mu)$}
 \\\hline\hline
\end{tabular}
\caption{The column $R^{(I,S)}_\mu(q,p)$ for isospin ($I$) and strangeness ($S$).}
\label{tab:states}
\end{table}

With the decomposition (\ref{def-proj}) the solution of the Bethe-Salpeter equation
takes the form
\begin{eqnarray}
&& T_{\mu \nu}(\bar k ,k ;w )  = \sum_{J,P}\,M^{(J,P)}(\sqrt{s}\,)\,
{\mathcal Y}^{(J,P)}_{\mu \nu}(\bar q, q;w) \,,
\nonumber\\
&& M^{(J,P)}(\sqrt{s}\,) = \Big[ 1- K^{(J,P)}(\sqrt{s}\,)\,J^{(J,P)}(\sqrt{s}\,)\Big]^{-1}\,
K^{(J,P)}(\sqrt{s}\,)\,,
\label{}
\end{eqnarray}
where the loop matrix $J^{(J,P)}_{ab}(\sqrt{s}\,)$ is diagonal in the coupled-channel space.
Diagonal elements are fully specified in terms of the meson and baryon-decuplet masses $m$ and $M$
entering the considered channel and a universal subtraction point $\mu=\mu(I,S)$ depending
only on the isospin and strangeness quantum numbers. For the loop functions of the
the different sectors with $J^P=(\frac{1}{2})^\pm,(\frac{3}{2})^\pm$ we find

\begin{table}
\tabcolsep=2.4mm
\begin{center}
%\begin{tabular}{||c||r|r|r|r|r|r|r|r|r|r||}
\begin{tabular}{||c||p{7mm}|p{7mm}|p{7mm}|p{7mm}|p{7mm}|p{7mm}|p{7mm}|p{7mm}|p{7mm}|p{7mm}||}
\hline
$(I, S)$            &11 &12 &22 &13 &23 &33 &14 &24 &34 &44 \\
 \hline\hline
($\frac12$, $-4$) &$-3$ &-- &-- &-- &-- &-- &-- &-- &-- & -- \\ %1
  \hline
(0,$-3$) &$\phantom{-}0$ &$-3$ &$\phantom{-}0$ &-- &-- &-- &-- &-- &-- & -- \\ %2
  \hline
(1,$-3$) &$\phantom{-}0$ &$-\sqrt 3$ &$-2$ &-- &-- &-- &-- &-- &-- & -- \\ %3
  \hline
($\frac12$, $-2$) &$\phantom{-}2$ &$-1$ &$\phantom{-}2$  &$\phantom{-}0$ &$-3$ &$\phantom{-}0$ & $\phantom{-}\frac{3}{\sqrt 2}$ &$\phantom{-}0$ &$\frac{3}{\sqrt 2}$ & $\phantom{-}3$ \\ %4
  \hline
($\frac32$, $-2$) &$-1$ &$\phantom{-}2$ &$-1$ &-- &-- &-- &-- &-- &-- & -- \\ %5
  \hline
($0$, $-1$) &$\phantom{-}4$ &$\sqrt 6$ &$\phantom{-}3$ &-- &-- &-- &-- &-- &-- & -- \\ %6
  \hline
($1$, $-1$) &$\phantom{-}2$ &$\phantom{-}1$ &$\phantom{-}4$ &$\phantom{-}0$ &$-\sqrt 6$ &$\phantom{-}0$ &$\phantom{-}2$ &$\phantom{-}0$ &$-\sqrt 6$ & $\phantom{-}1$ \\ %7
  \hline
($2$, $-1$) &$-2$ &$\sqrt 3$ &$\phantom{-}0$ &-- &-- &-- &-- &-- &-- & -- \\ %8
  \hline
($\frac12$, $0$) &$\phantom{-}5$ &$-2$ &$\phantom{-}2$ &-- &-- &-- &-- &-- &-- & -- \\ %9
  \hline
($\frac32$, $0$) &$\phantom{-}2$ &$\phantom{-}0$ & $\phantom{-}0$ &$\sqrt{\frac52}$ & $\phantom{-}\frac{3}{\sqrt 2}$ & $-1$ &-- &-- &-- & -- \\ %10
  \hline
($\frac52$, $0$) & $-3$ &-- &-- &-- &-- &-- &-- &-- &-- & -- \\ %11
  \hline
($1$, $1$) & $\phantom{-}1$ &-- &-- &-- &-- &-- &-- &-- &-- & -- \\ %12
   \hline
($2$, $1$) & $-3$ &-- &-- &-- &-- &-- &-- &-- &-- & -- \\ %13
\hline
\end{tabular}
\caption{The coefficient matrices $C^{(I,S)}$ of the Weinberg-Tomozawa term that
characterize the meson baryon-decuplet interaction introduced in
(\ref{def-c}). The  matrix elements are ordered according to the states in
Tab.~\ref{tab:states}.} \label{tab:coeff}
\end{center}
\end{table}

\begin{eqnarray}
&& J^{(J,P)}(\sqrt{s}\,)= N^{(J,P)}(\sqrt{s}\,)\,\Big(I(\sqrt{s}\,) - I(\mu) \Big)\,,
\nonumber\\
&& I(\sqrt{s}\,)=\frac{1}{16\,\pi^2}
\left( \frac{p_{\rm cm}}{\sqrt{s}}\,
\left( \ln \left(1-\frac{s-2\,p_{\rm cm}\,\sqrt{s}}{m^2+M^2} \right)
-\ln \left(1-\frac{s+2\,p_{\rm cm}\sqrt{s}}{m^2+M^2} \right)\right)
\right.
\nonumber\\
&&\qquad \qquad + \left.
\left(\frac{1}{2}\,\frac{m^2+M^2}{m^2-M^2}
-\frac{m^2-M^2}{2\,s}
\right)
\,\ln \left( \frac{m^2}{M^2}\right) +1 \right)+I(0)\;,
\label{i-def}
\end{eqnarray}
where $\sqrt{s}= \sqrt{M^2+p_{\rm cm}^2}+ \sqrt{m^2+p_{\rm cm}^2}$ and $E=\sqrt{M^2+p_{\rm cm}^2}$ and
\begin{eqnarray}
&& N^{(1/2,\pm)}(\sqrt{s}\,) = \Big(E \mp M \Big)\,
\left(\frac{2}{9}\,\frac{E^2}{M^2}-\frac{2}{9}\, \right)\,,
\nonumber\\
&& N^{(3/2,\pm)}(\sqrt{s}\,) = \Big(E \mp M \Big)\, \left(
\frac{5}{9}+\frac{2}{9}\,\frac{E}{M}+\frac{2}{9}\,\frac{E^2}{M^2}\right)\,.
\label{def-norm}
\end{eqnarray}

It is left to identify the interaction kernel $K^{(J,P,I,S)}_{ab}(\sqrt{s}\,)$
as predicted by the Weinberg-Tomozawa term (\ref{WT-term}). The coupled-channel
structure is made explicit, with $a$ and $b$ referring to initial and final states,
\begin{eqnarray}
&& K^{(\frac{1}{2},\pm ,I,S)}_{ab}(\sqrt{s}) =
\frac{C^{(I,S)}_{ab}}{4\,f^2}\,\Big( 2\,\sqrt{s}\mp M_a \mp M_b\Big) \,,
\nonumber\\
&& K^{(\frac{3}{2},\pm ,I,S)}_{ab}(\sqrt{s}) = \frac{C^{(I,S)}_{ab}}{4\,f^2}\,\Big( 2\,\sqrt{s}\pm M_a \pm M_b\Big) \,,
\label{def-c}
\end{eqnarray}
and $M_a$ and $M_b$ denoting the isospin averaged decuplet masses of initial and final channel.
The interaction kernels are identical for the two channels, $J=\frac{1}{2}$ and $J=\frac{3}{2}$,
considered here. The $'\pm'$ in (\ref{def-c}) keeps track of the parity quantum number with $P=\pm\,1$.
The values for the coefficient matrix $C^{(I,S)}_{ab}$ characterizing the strength of the
Weinberg-Tomozawa term in the coupled-channel space are collected in Tab.~\ref{tab:coeff}.
A remark of caution is in order here. Though we provide here the contribution of the
Weinberg-Tomozawa term to the leading channels, it should be
emphasized that only for $J^P=(\frac{3}{2})^- $ the term constitutes the dominant
contribution to the interaction kernel. In the remaining channels there exist further
terms in the chiral Lagrangian, so far basically unknown, that may change the interaction
kernel significantly. Nevertheless it is instructive to study the consequence of the
Weinberg-Tomozawa term in those channels as it is. One may draw
qualitative conclusion as to whether the chiral Lagrangian has the potential to dynamically
generate $J^P=(\frac{3}{2})^+ $ and $J^{P}=(\frac{1}{2})^\pm$ baryon resonances as well.

\section{Results}

In order to study the formation of baryon resonances we generate speed plots as suggested
by H\"ohler \cite{Hoehler:speed}. The speed, ${\rm Speed}_{ab}^{(J,P)}(\sqrt{s})$, of a given
channel $a\,b$ is introduced by \cite{Hoehler:speed,speed},
\begin{eqnarray}
&& t^{(J,P)}_{ab}(\sqrt{s}\,)=\frac{1}{8\,\pi \,\sqrt{s}}\,
\Big( p^{(a)}_{\rm cm}\,N_a^{(J,P)}(\sqrt{s}\,)\,p_{\rm cm}^{(b)}\,
N_b^{(J,P)}(\sqrt{s}\,)\Big)^{1/2}\,M_{ab}^{(J,P)}(\sqrt{s}\,)\,,
\nonumber\\
&& {\rm Speed}_{ab}^{(J,P)}(\sqrt{s}\,) = \Big|\sum_{c}\,
 \Big[\frac{d}{d \,\sqrt{s}}\, t_{ac}^{(J,P)}(\sqrt{s}\,)\Big]\,
 \Big(\delta_{cb}+2\,i\,t_{cb}^{(J,P)}(\sqrt{s}) \Big)^\dagger
\Big| \,.
\label{}
\end{eqnarray}
If a resonance with not too large decay width sits in the amplitude $M^{(J,P)}(\sqrt{s})$ a
clear peak structure emerges in the speed plot even if the resonance structure
is masked by a background phase.

We begin with a discussion of our results in the SU(3) limit. The latter is
not defined uniquely depending on the magnitude of the current quark masses,
$m_u=m_d=m_s$. We study this dependence at chiral order $Q^2$. Using the meson-baryon sigma
terms of \cite{LK02} a meson mass $m_{[8]} \simeq m_K$ implies for instance a baryon-octet mass
$M_{[8]}\simeq 1275 $ MeV. Applying the large-$N_c$ limit one may use $M_{[8]}=M_{[10]}$ for the
baryon-decuplet masses. This scenario we call the 'heavy' SU(3) limit. Similarly we introduce the
notion of a 'light' SU(3) limit with $m_{[8]}= m_\pi$ and $M_{[8]} \simeq 860$ MeV.

\begin{figure}[t]
\begin{center}
\includegraphics[width=14.0cm,clip=true]{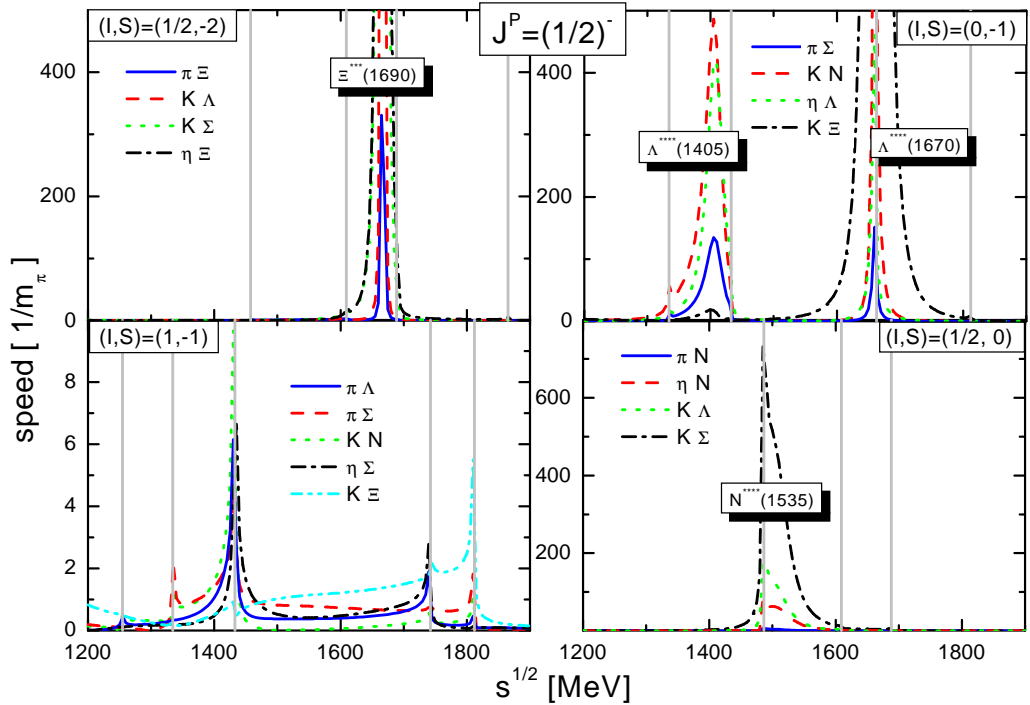}
\end{center}
\caption{Diagonal speed plots of the $J^P=(\frac{1}{2})^-$ sector. The vertical lines show the
opening of inelastic meson baryon-decuplet channels. Parameter-free results are obtained
in terms of physical masses and $f=90$ MeV \cite{Granada}.} \label{fig2}
\end{figure}

Before discussing our results for meson baryon-decuplet scattering we briefly recall the results
of the recent work \cite{Granada} which addressed the chiral dynamics of s-wave
meson baryon-octet states. In the SU(3) limit meson baryon-octet scattering is classified
according to,
\begin{eqnarray}
8 \otimes 8 = 27 \oplus \overline{10} \oplus 10 \oplus 8 \oplus 8 \oplus 1 \,.
\label{oc-deco}
\end{eqnarray}
The Weinberg and Tomozawa interaction predicts attraction in the two 8-plet and the 1-plet channel but repulsion
in the 27-plet channel \cite{Jido03}. The repulsion in the 27-plet channels follows for instance
from the negative signs of the coefficients $C^{(I,S)}$ in the (1,1) and (2,-1) sectors that
involve one channel only (see e.g. \cite{LK02}). All together the 27-plet has contributions in
nine channels with
$$(I,S)_{[27]}=(1,1),({\textstyle{3\over 2}},0),({\textstyle{1\over 2}},0),(2,-1),(1,-1),(0,-1),
({\textstyle{3\over 2}},-2),({\textstyle{1\over 2}},-2),(1,-3)\,.$$
Similarly the $\overline{10}$-plet and 10-plet contribute in
$$(I,S)_{\overline{10}}=(0,1),({\textstyle{1\over 2}},0),(1,-1),({\textstyle{3\over 2}},-2)\,,$$
$$\;\;(I,S)_{10}=({\textstyle{3\over 2}},0),(1,-1),({\textstyle{3\over 2}},-2),(0,-3)\,.$$
The vanishing of $C^{(0,1)}$
(see e.g. \cite{LK02}) tells that the interaction strength
vanishes in the $\overline{10}$-plet channel \cite{Jido03}. An analogous result \cite{Jido03}
holds for the 10-plet channel. This follows most efficiently from the vanishing of $C^{(0,-3)}$.
As a consequence in the 'heavy' SU(3) limit the chiral dynamics predicts two degenerate
octet bound states together with a non-degenerate singlet state
\cite{Wyld,Dalitz,Ball,Rajasekaran,Wyld2,Granada,Jido03}.
In the 'light' SU(3) limit all states disappear leaving no clear signal in any of the
speed plots. Using physical meson and baryon-octet masses the bound-state turn into
resonances as shown in Fig. \ref{fig2}. The speed plots show strong evidence for the formation of
the $\Xi(1690)$, $\Lambda(1405)$, $\Lambda(1670)$ and $N(1535)$ resonances. The 'disappearance'
of the remaining states was discussed in detail
in \cite{Granada}.

We turn to the meson baryon-decuplet scattering process, --- the focus of this work.
In the SU(3) limit this  process decouples into four different channels, labelled
according to,
\begin{eqnarray}
8 \otimes 10 = 35 \oplus 27 \oplus 10 \oplus 8\,.
\label{8-10-decom}
\end{eqnarray}
In the $J^P=\frac{3}{2}^-$ sector the Weinberg-Tomozawa interaction is attractive in
the 8-plet, 10-plet and 27-plet channel, but repulsive in the 35-plet channel. Therefore one
may expect resonances or bound states in the former channels. For instance,
the repulsion in the 35-plet channel with
\begin{eqnarray}
(I,S)_{35} &=& (2,1),({\textstyle{5\over 2}},0),({\textstyle{3\over 2}},0),(2,-1),(1,-1),
({\textstyle{3\over 2}},-2), ({\textstyle{1\over 2}},-2), \nonumber\\
&&(1,-3),(0,-3),
({\textstyle{1\over 2}},-4)
\,,
\end{eqnarray}
follows from the negative $C^{(I,S)}$ coefficients
(see Tab. \ref{tab:coeff}) in the $(2,1)$ and $(\frac{1}{2},-4)$ sectors. Similarly the
positive $C^{(1,1)}$ coefficients reflects the weak attraction in the 27-plet channel.
Indeed, in the 'heavy' SU(3) limit we find $72=4\times (8+10)$ bound states of masses
$ 1620$ MeV and $ 1710$ MeV forming an octet and decuplet representation of the SU(3) group.
We do not find a 27-plet bound state reflecting the weaker attraction in this channel.
However, if we artificially increase the amount of attraction by about only 20 $\%$
by lowering the value of $f$ a clear bound state arises in this
channel. These phenomena persist if we use somewhat larger baryon-decuplet masses. A contrasted
result is obtained if we lower the meson masses down to the pion mass arriving at
the 'light' SU(3) limit. Then we find neither bound nor resonance
octet or decuplet states. Increasing the baryon-decuplet mass somewhat away from
the baryon-octet mass does not change this result.

Even though the Weinberg-Tomozawa term does not constitute the complete leading order
interaction in the $J^P=(\frac{1}{2})^\pm$ and $J^P=(\frac{3}{2})^+$ channels it is nevertheless
instructive to explore the consequences thereof in those sectors. Using physical masses
we observe  in the speed plots of the $J^P=(\frac{1}{2})^+$ sector octet and decuplet resonances
with masses centering around 2000 MeV. Of course the resonances of  this channel are
dominated by s-wave meson baryon-octet channels \cite{Granada} and one should not expect any
realistic results in terms of d-wave meson baryon-decuplet channels only. Similarly
we do not find any clear signal of $J^P=(\frac{1}{2})^-$ and $J^P=(\frac{3}{2})^+$
resonances in the p-wave channels of masses below 2000 MeV.

\begin{figure}[t]
\begin{center}
\includegraphics[width=14.0cm,clip=true]{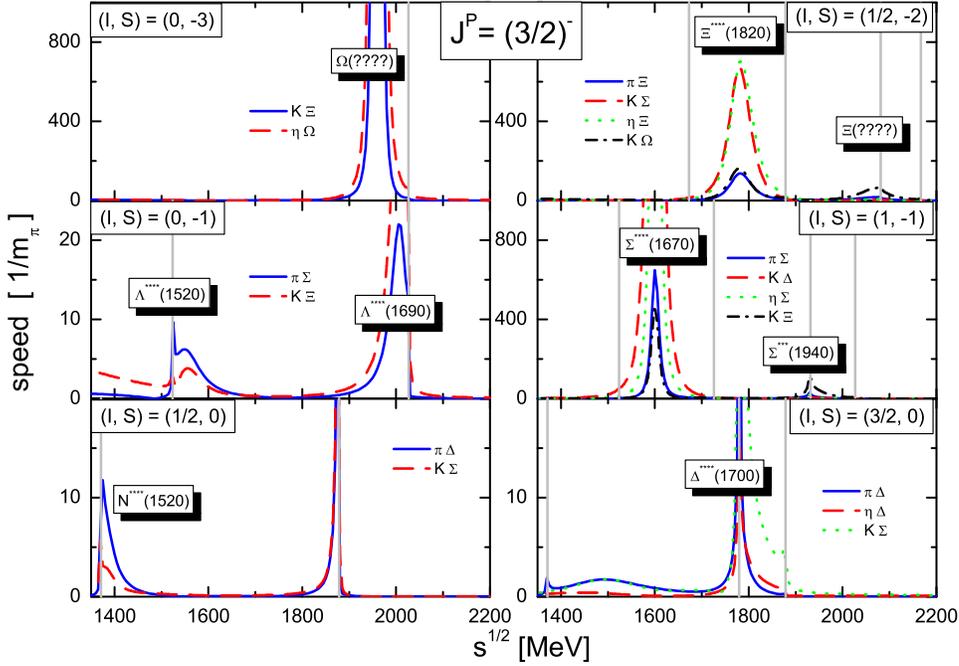}
\end{center}
\caption{Diagonal speed plots of the $J^P=(\frac{3}{2})^-$ sector. The vertical lines show the
opening of inelastic meson baryon-decuplet channels. Parameter-free results are obtained
in terms of physical masses and $f=90$ MeV.} \label{fig1}
\end{figure}

We turn to the central result of this work. In Fig. \ref{fig1} speed plots of the $J^P=(\frac{3}{2})^-$
sector are shown for all channels in which octet and decuplet resonance states are expected.
It is a remarkable success of the present scheme that it predicts parameter free the four-star
hyperon resonances $\Xi(1820)$, $\Lambda(1520)$, $\Sigma (1670)$ with masses quite close to the
empirical values. The nucleon and isobar resonances $N(1520)$ and $\Delta (1700)$ also
present in Fig. 1, are predicted with less accuracy. The important result here is the fact
that those resonances are generated at all. It should not be expected to arrive already at
fully realistic results in this leading order calculation. For instance chiral correction
terms provide a d-wave $\pi \,\Delta$-component of the $N(1520)$.

We continue with the peak in the (0,-3)-speeds at mass 1950 MeV.
Since this is below all thresholds it is in fact a bound state. Such a state
has so far not been observed but is associated with a decuplet resonance.
It is a clear and solid prediction of our scheme and we suggest to search for such a state.
Typically such a state is predicted to have resonance character in
large-$N_c$ phenomenology and quark-model calculations with a mass above 2000 MeV \cite{Schat}.
Further states belonging to the decuplet are seen in the $(\frac{1}{2},-2)$- and
$(1,-1)$-speeds at masses 2100 MeV and 1920 MeV. The latter state can be identified with the
three star $\Xi (1940)$ resonance. Finally we point at the fact that the $(0,-1)$-speeds show
signals of two resonance states consistent with the existence of the four star resonance
$\Lambda(1520)$ and $\Lambda(1690)$ even though in the 'heavy' SU(3) limit we observed only
one bound state. It appears that the SU(3) symmetry breaking pattern generates the 'missing'
'singlet' state not predicted by (\ref{8-10-decom}).

It should be noted that the mass and widths parameters one may extract from the speed plots in Fig. 1
could be improved by incorporating contributions from meson baryon-octet
decays not considered here. In particular one would expect from previous
phenomenological studies like \cite{LWF02} that the vector-meson baryon-octet channels
may play an important role in the course of improving the results of this work.
One may speculate that the weak attraction found in the 27-plet channel which is almost
strong enough to generate resonance structures could generate
states with anomalous quantum numbers like $(I,S)=$(1,1) if the correction terms
conspire to slightly increase the attraction found already at leading order.

{\bfseries{Acknowledgments}}

The authors thank the ECT$^*$ for hospitality and pleasant working conditions.
E.K. acknowledges partial support from GSI.

%%%%%%%%%%%%%%%%%%%%%%%%%%%%%%%%%%%%%%%%%%%%%%%%%%%%%%%%%%%%%%%%%%%%%

\end{document}